\documentclass[authordraft, nonacm, review=false, timestamp=false]{acmart}

\AtBeginDocument{%
  }

\usepackage{xurl}
\usepackage{hyperref}
\usepackage{xspace}
\usepackage{todonotes}

\begin{document}

\title[Digital Sovereignty, Digital Patriotism, and the Emerging Geopolitics of Software Adoption]{Searching for European Alternatives: Digital Sovereignty, Digital Patriotism, and the Emerging Geopolitics of Software Adoption}

\author{Advait Sarkar}
\email{advait.sarkar@cl.cam.ac.uk}
\affiliation{%
  \institution{University of Cambridge}
  \city{Cambridge}
  \country{United Kingdom}
}

\affiliation{%
  \institution{University College London}
  \city{London}
  \country{United Kingdom}
}

\begin{abstract}
Software adoption has traditionally been understood through instrumental lenses, such as usability, cost, security, and interoperability. We argue that a new, ideological dimension is reshaping adoption decisions: one we term digital patriotism, the individual counterpart to the state ideology of digital sovereignty. Through two studies, we trace this phenomenon. First, a directed content analysis of decisions made by European government agencies to switch away from de facto technology standards reveals a shift around 2020: early switches cited costs and vendor lock-in, while later switches invoke sovereignty, geopolitical risk, and investment in local industry. Second, a qualitative analysis of over 700 online comments (over 51,000 words) surfaces how consumers and businesses articulate motivations for seeking European software alternatives. We find that digital patriotism entails a willingness to accept functional compromise in service of ideological goals. Our work extends software adoption theory by drawing attention to value rationality alongside instrumental rationality, and contributes an empirical account of how geopolitics is reshaping technology choice in the workplace.
\end{abstract}

\ccsdesc[500]{Human-centered computing~Human computer interaction (HCI)}
\ccsdesc[300]{Human-centered computing~HCI theory, concepts and models}
\ccsdesc[300]{Human-centered computing~Empirical studies in HCI}
\ccsdesc[500]{Social and professional topics~Computing / technology policy}
\ccsdesc[500]{Social and professional topics~Government technology policy}
\ccsdesc[300]{Social and professional topics~Governmental regulations}

\keywords{Value rationality, technology nationalism, platform governance, open-source software, public sector IT, online discourse analysis, techno-politics}

\maketitle

\newcommand{\pquote}[2]{\textcolor{darkgray}{\emph{``#2''}}~(U#1)}
\newcommand{\el}{[\dots]\xspace}

\section{Introduction}
Reflect for a moment on how you came to choose the software you use: your operating system, your search engine, your word processor. Across Europe, a question is being asked in city councils, office boardrooms, and online forums alike: \emph{Is there a European alternative?}

Not long ago, this question would have seemed irrelevant. Conventional accounts of why people choose to use the software they do (detailed in Section~\ref{sec:background-software-adoption}), particularly in the workplace, focus largely on instrumental concerns: how much does the software cost? How interoperable is the software with my other tools and file formats? Does it have more features than its competitors? How easy to use is it? Accounts of software adoption typically do not consider how software choice is a mode of ideological and political expression (though the fact that software is political has certainly not escaped notice in prior research, detailed in Section~\ref{sec:background-hci-politics}).

In this paper, we investigate how one geopolitical current is becoming an increasingly influential force for software choice in Europe. In the years 2020-2025, a confluence of political events (pandemic-era supply chain anxieties, the war in Ukraine, deteriorating transatlantic relations) brought to the forefront the European ideology of \textbf{digital sovereignty} (detailed in Section~\ref{sec:background-digital-sovereignty}), namely, the desire for the State to have sovereign control over technology and the power exercised through it. Through two studies, we show how this ideology is now translating into software adoption decisions. European governments, businesses, and individuals are now consciously making software choices motivated by this ideology.

The broad research question that motivates us is \emph{how does the concept of digital sovereignty in Europe manifest in software adoption choices?} Since this is a state ideology, one obvious starting point is to look at how states themselves choose software. We first present an analysis of ``switching'' events in European governments: documented instances where a government body decided to switch away from de facto standards in one or more software categories (Section~\ref{sec:study1}). We use the term \textbf{de facto standards} to mean market leaders in their corresponding software categories, such as search engines, operating systems, cloud providers, email providers, etc. We investigated switching events through directed content analysis of a heterogeneous set of sources including press releases and government websites. We examined motivations for switching, and what alternatives were proposed. We find that prior to 2020, switching events were motivated by instrumental concerns such as reducing costs and avoiding vendor lock-in, and the solution proposed was a coordinated use of open-source software. However, after 2020, switching events are motivated by the ideology of digital sovereignty, and the solutions proposed expand to include developing state-owned alternatives and investment in the domestic private sector.

However, software adoption choices at the state level are not necessarily representative of everyday consumer and business choices. We therefore further ask, \emph{how can digital sovereignty manifest in consumer and business choices?} If sovereignty is an ideal that only operates at the state level, what corresponding ideal might be at play at the individual level? To begin to address these questions, we present an analysis of over 700 comments on the online forum Hacker News, primarily from software developers and business owners, in threads discussing switching decisions (Section~\ref{sec:study2}). We find that while some instrumental motivations for switching remain (e.g., the de facto standards don't meet certain functional requirements), commenters cite a variety of ideological motivations that come together in a phenomenon we term \textbf{digital patriotism}. In our dataset, we see that these motivations lead European software consumers to search specifically for European alternatives to the de facto standards, despite challenges and limitations encountered.

In closing, we discuss how the state ideology of digital sovereignty may translate into the individual ideology of digital patriotism through community efforts, such as community fora and catalogues of European alternatives to de facto standards (Section~\ref{sec:discussion-catalogues}), and how the situation is still evolving as de facto standards adapt their products to be viewed as consistent with the ideology of sovereignty, and European governments view co-operation with de facto standards as part of their sovereignty strategy (Section~\ref{sec:discussion-adaptation}).

\section{Background and Related Work}
\label{sec:background}

\subsection{Digital Sovereignty}
\label{sec:background-digital-sovereignty}

Digital sovereignty in the European context has been defined as \textit{``Europe's ability to act independently in the digital world''} \cite{madiega2020digital_sovereignty_europe}. Motivated by the observation that the \textit{``dependence on foreign technology [... presents] a risk to Europe's influence''}, the concept of digital sovereignty has gained currency in recent years, and is invoked as motivation for a number of significant European policy measures to \textit{``narrow the investment gap''}, \textit{``adapt EU industrial and technological 
capacity to the competitive environment''}, help \textit{`` European citizens \el recover control of their digital data''}, and for \textit{``rethinking the governance mechanisms currently operating within the EU''} \cite{madiega2020digital_sovereignty_europe}.

The concept emerged in the late 1990s but gained prominence after key catalytic events. The 2013 Snowden revelations dramatically increased perceptions of US digital hegemony \cite{fratini2024digital, santaniello2025attributes}. In 2016, the EU's GDPR, China's Cybersecurity Law, and American protectionist policies catalysed global discourses \cite{pohle_thiel_digital_2020}. France developed ``souveraineté numérique'' in opposition to US tech dominance and Germany adopted a similar approach post-Snowden \cite{pohle_thiel_digital_2020, MaC821}. In Europe the February 2020 Commission digital strategy first introduced and defined an explicit notion of technological sovereignty: \textit{``ensuring the integrity and resilience of our data infrastructure, networks and communications \el creating the right conditions for Europe to develop and deploy its own key capacities, thereby reducing our dependency on other parts of the globe for the most crucial technologies [... and reinforcing] Europe’s ability to define its own rules and values''}  \cite{europeancommission2020_shaping_digital_future}, and shortly thereafter the discourse assumes the more familiar phraseology of ``digital sovereignty'' in the July 2020 EPRS briefing ``Digital sovereignty for Europe'' \cite{madiega2020digital_sovereignty_europe}.

As one might imagine, digital sovereignty is a contested concept with multiple definitions across disciplines. This ambiguity enables strategic deployment by diverse actors \cite{pohle2024unthinking}. Fratini et al. \cite{fratini2024digital} identify four models: rights-based, market-oriented, centralisation, and state-based, each pursuing distinct values through different policy instruments. Santaniello \cite{santaniello2025attributes} proposes a framework structured around five attributes: adversariality, multiversity, latency, instrumentality, and hypocrisy. Hummel et al. \cite{hummel2021data} relate digital sovereignty to terms such as data sovereignty and cyber sovereignty, noting their overlapping use. Pohle and Thiel \cite{pohle_thiel_digital_2020} distinguish institutionalist approaches emphasising state authority from constructivist perspectives viewing sovereignty as discursive resource. Floridi \cite{floridi2020fight} emphasises its particular importance for the EU's geopolitical positioning.

Digital sovereignty manifests through diverse policy instruments. Data localisation (where data is physically stored) is one focus of legislation \cite{christakis2023european}. Roberts et al. \cite{roberts2021safeguarding} reviewed several EU policy measures for digital sovereignty across data governance (e.g., GDPR), constraining platform power (e.g., the Digital Markets Act), digital infrastructures, emerging technologies, and cybersecurity. China, which formally articulated cyber sovereignty in 2010 \cite{hung2025exploring}, pursues territorialisation of infrastructure, indigenisation of technology, and comprehensive content control through this doctrine \cite{creemers2020china}. China's model has been interpreted as a ``norm subsidiary response'' \cite{hung2025exploring}, which explains how a relatively weaker state can challenge dominant norms, presenting an alternative to Western models, and contributing to what some term a ``Digital Cold War'' \cite{gao2022attractive}. The EU reponse may similarly exercise regulatory power through the ``Brussels Effect,'' where its standards become global norms \cite{bradford2020brussels}.

There are critiques as well as alternative deployments of the concept of sovereignty. A notable example is how indigenous communities have claimed sovereignty over their data to resist extractivism \cite{walter2020indigenous}. Mueller \cite{mueller2020against} critiques sovereignty's application to cyberspace as inappropriate, threatening transnational governance and risking the Balkanisation of the internet. Farrand \& Carrapico \cite{farrand2022digital} identify "regulatory mercantilism" in EU policies that favour domestic actors while framing foreign companies as threats. Similarly, Calvo et al. \cite{calvo2025responding} analyse how different governments have responded to the political power of ``platform firms'' (loosely corresponding to what we call in this paper de facto standards), finding that diverging national priorities are inducing \textit{``broader changes to the structure of what was [...] a relatively borderless global online economy''}.

Despite extensive prior research on the differing conceptualisations of digital sovereignty, their history, and their policy implications, little attention has been paid to how this ideology concretely affects software adoption choices in the workplace.

\subsection{Software Adoption Models}
\label{sec:background-software-adoption}

The Diffusion of Innovations theory \cite{Rogers2003-qu} was an early (and still influential) account of how innovations spread through social systems via five attributes affecting adoption rate: relative advantage (perceived improvement over predecessors), compatibility (alignment with existing values and practices), complexity (perceived difficulty, inversely related to adoption), trialability (ability to test with limited commitment), and observability (visibility of benefits to others). 

Similarly, the Technology Acceptance Model (TAM) \cite{davis1989perceived}, posits that perceived usefulness and perceived ease of use determine user attitudes and behavioral intentions toward information systems. Building on the TAM and other models is the Unified Theory of Acceptance and Use of Technology (UTAUT) \cite{venkatesh2003user}, identifying four core determinants: performance expectancy, effort expectancy, social influence, and facilitating conditions. Its extension, UTAUT2 \cite{venkatesh2012consumer}, adds hedonic motivation, price value, and habit for consumer contexts. Hedonic motivation, defined as the ``fun or pleasure derived from using technology'' \cite{tamilmani2019battle} is an interesting factor from our perspective, as it is a non-instrumental determinant of software adoption. However, it cannot be characterised as a coherent value or ideology.

The Technology-Organisation-Environment framework \cite{depietro1990context} addresses organisational adoption through three contexts: technological (ICT infrastructure, relative advantage), organisational (firm size, management support, resources), and environmental (competitive pressure, regulations, trading partners). The Task-Technology Fit model \cite{goodhue1995task} emphasizes alignment between technology characteristics and task requirements. The Model of PC Utilisation \cite{thompson1991personal} identifies job-fit, complexity, long-term consequences, affect, facilitating conditions, and social factors as determinants. While ``social factors'' might seem a relevant prior for our interest in ideology-driven software adoption, in this case it is simply operationalised as the proportion of co-workers who use the technology and how supportive the organisation is of using the technology.

Studies of user resistance to software adoption have shown that users may resist new systems when they perceive threats to their autonomy, professional identity, or ethical standards. This resistance is particularly pronounced in contexts where software is introduced without adequate consideration of existing practices \cite{laumer2011people, lin2018user}.

Previous work has thus characterised software adoption as depending on demonstrable benefits, usability, compatibility, trialability, cost-effectiveness, organizational support, and social endorsement, while complexity, incompatibility, inadequate support, violations of autonomy, and prohibitive costs impede diffusion. The literature on software adoption and diffusion is thus largely silent when it comes to the role of ideology, political or otherwise.

\subsection{Political Values in Software}
\label{sec:background-hci-politics}

It is obligatory in any discussion of the politics of technology to acknowledge Winner \cite{winner1980artifacts}, who argued that technology can embody politics either by requiring specific social arrangements or by being designed in a way that results in settling political issues. A philosophy of interaction design that ensues is value sensitive design \cite{friedman2019value}, a method for designers to embed their values into the design of technology. Values in software can be ``techno-generic'' (inherent to the technology) or ``domain-specific'' (rooted in social contexts) \cite{Gogoll2026values}.

The ideological dimensions of open source and free software movements have also been studied. Free software developers often claim political neutrality while simultaneously enacting deeply political values such as transparency, freedom, and meritocracy \cite{coleman2012coding}. Coleman's concept of the ``inadvertent politics of contrast'' \cite{coleman2004political} captures how free software becomes political by virtue of its opposition to proprietary models. Similarly, Kelty introduces the idea of ``recursive publics'' \cite{kelty2008two}, which can be crudely defined as communities that create and maintain the very tools that enable their collective existence.

The ideology of open source and free software might at first glance appear to support digital sovereignty, yet in practice (and as we will see in our data) they are uneasy bedfellows. Pannier observes that \textit{``government involvement in open source is not only pragmatic; it is increasingly politicized, and serves to uphold governments’ ambitions for national security, international influence, or digital sovereignty''}, and yet there are \textit{``tensions between the desire to secure universally used, critical open source components, the desire to develop “sovereign” technologies, and the risk of encroaching on the horizontal and decentralized functioning of open source''} \cite{pannier_software_2022}.

Despite the deeply ideological nature of the open source software movement (a recent review is given by Yue et al. \cite{yue2023towards}), prior studies have described a lack of strong evidence that open source adoption in organisations is systematically driven by ideology \cite{ven2008impact}. In light of a rapidly evolving geopolitical landscape and the rapidly intensifying focus on digital sovereignty, our study revisits this issue and contributes a new empirical account of how workplace software adoption decisions are increasingly ideologically driven.

\section{Study 1: The Metamorphosis of Technology Procurement Rationales in European Governments}
\label{sec:study1}

To understand how the technology procurement rationales of European governments have evolved over time, we conducted a directed content analysis \cite{hsieh2005three, assarroudi2018directed} of grey literature documenting high profile ``switching events''; namely, events where a public sector body had announced the decision to transition some aspect of their technology use away from de facto standards.

\subsection{Method}

Due to the extreme heterogeneity of these events and the associated documentation, a wide set of online sources were compiled using a combination of ad hoc web search and searches on specific websites that document such events, such as the website of the Free Software Foundation Europe, ZDNet, The Register, the Interoperable Europe Portal, and The Document Foundation. Searches were repeated over a period of 6 months from June to December 2025. Sources were further accumulated through ``snowballing''; press reporting on switching events often cited and linked to other prior switching events, which were investigated. Wherever possible, efforts were made to trace press coverage to official government documents or authorised first party press releases.

A formalised sampling frame is neither feasible nor epistemically appropriate because the putative population of European technology ``switch'' decisions is not systematic across jurisdictions and languages, much of the documentary evidence sits in grey or administrative sources, and what becomes visible is affected by media coverage. Our method can be described as purposeful sampling of critical and maximum‑variation cases \cite{palinkas2015purposeful}, aiming for analytic, not statistical generalisation.

The full set of sources for switching events is presented in Appendix~\ref{apx:government-switches}. In the following section this appendix is referenced using the notation [S-N] to disambiguate these references from standard citations.

The collected sources were subjected to a highly directed content analysis aimed at two questions: what were the stated motivations for switching, and what the proposed solution was, i.e., what was being switched \emph{to}.

\subsection{Results}

\begin{table*}[!t]
  \caption{Government switches. The Source ID column refers to IDs in Appendix~\ref{apx:government-switches}.}
  \label{tab:government-switches}
  \footnotesize
  \begin{tabular}{c p{3.5cm} p{3.5cm} p{3cm} c}
    \toprule
    Year & Event & Motivations & Solution & Source ID\\
    \midrule
    2005 & French Gendarmerie completes switch to OpenOffice & Cost & OpenOffice & S-1\\
    2006 & French National Assembly confirms move to open source & Cost & OpenOffice & S-2\\
    2008 & French Gendarmerie begins GendBuntu project & Flexibility, cost & Develop own OS, switch to other OSS & S-3\\
    2011 & Copenhagen hospitals announce move to LibreOffice in 2012 & Cost, customer support & LibreOffice & S-4\\
    2012 & Leipzig, Germany switches to OpenOffice and LibreOffice & Lock-in & Various OSS solutions & S-5\\
    2012 & Toulouse switches to LibreOffice & Lock-in & Various OSS solutions & S-6\\
    2012 & Greek municipality of Pilea-Hortiatis migrates to LibreOffice & Cost & LibreOffice & S-7\\
    2013 & South Tyrol, Italy, switches to LibreOffice & Cost & LibreOffice & S-8\\
    2013 & Valencia Generalitat completes deployment of free office software & Cost, promoting local ICT industry, promoting Valencian language, localisation, competition & LibreOffice, LliureX operating system based on Linux & S-9\\
    2014 & LibreUmbria - Umbria Region, Italy, migrates to LibreOffice & Cost & LibreOffice & S-10\\
    2014 & UK Government adopts open document formats & Cost & ODF, PDF/A, and HTML & S-11\\
    2015 & Italian Ministry of Defence commits to LibreOffice and ODF & Independence from proprietary software & LibreOffice, ODF & S-12\\
    2016 & Barcelona City Council Digital Plan released & Digital sovereignty, lock-in & Data governance, custom OS built on free/OSS & S-13\\
    2017 & City of Barcelona migrates away from de facto standard operating system & Cost, promoting local ICT & Linux, OSS office, coordinated open source & S-14\\
    2018 & Tirana, Albania switches to LibreOffice & Lock-in & LibreOffice & S-15\\
    2020 & Schleswig-Holstein report on open source software & Cost, digital sovereignty & Open source & S-16\\
    2021 & Dortmund, Germany commits to using open source where possible & Efficiency, public money public code & Open source & S-17\\
    2024 & Schleswig-Holstein announces digitally sovereign IT workplace & Digital sovereignty, control over data, high costs of de facto standards, invest in local economy & Switch to open source for Office, OS, Cloud, telephony, etc. & S-18\\
    2025 & Dutch parliament passes 8 motions to support government switch & Digital sovereignty, local investment & Invest in private sector and nationalised home-grown alternatives & S-19\\
    2025 & Copenhagen and Aarhus municipalities announce phase out of de facto standard & Digital sovereignty, uncertainty under American administration, costs & European private sector alternatives & S-20\\
    2025 & Denmark Ministry of Digital Affairs moves away from de facto standard & Digital sovereignty, risk, freedom & LibreOffice & S-21\\
    2025 & Lyon announces Open Digital Territory, migrates from de facto standard & Digital sovereignty, environment & Coordinated use of OSS & S-22\\
    \bottomrule
  \end{tabular}
\end{table*}

A table of switching events in chronological order, with their stated motivations for switching and proposed solutions, is given in Table~\ref{tab:government-switches}. The analysis reveals a shift in motivations and solutions around the year 2020.

Switching events begin as early as 2005, with the French Gendarmerie switching to OpenOffice, citing the high costs of the de facto standards [S-1], followed by the French National Assembly switching to open source alternatives for their operating system, web browser, office productivity software, and email client in 2006, again citing cost savings [S-2]. In 2008, the Gendarmerie began the GendBuntu project, a bid to develop an Operating System and transition to more Open Source Software (OSS), citing greater flexibility and costs [S-3]. 

In 2011, a group of public hospitals in Copenhagen announced the transition to LibreOffice, citing the high costs and poor customer support of the defacto standards [S-4]. The transition of office productivity software towards free and open-source alternatives, such as LibreOffice and OpenOffice, is a recurring theme in early switches. Examples of such switches include the city of Leipzig in Germany in 2012, citing the avoidance of vendor lock-in as a motivation [S-5], Toulouse in France in 2012, also citing lock-in [S-6], the Greek municipality of Pilea-Hortiatis in 2012, citing costs [S-7], South Tyrol in Italy in 2013, citing costs in the wake of Italian austerity measures [S-8], the Umbria region of Italy in 2014 citing costs [S-10], and Tirana in Albania in 2018, citing vendor lock-in [S-15]. 

Wider moves to the Open Document Format (ODF), which facilitates interoperability between proprietary and OSS office productivity software, were also part of the story. In 2014, the UK Government announced the adoption of open document formats citing cost concerns [S-11], and in 2015 the Italian Ministry of Defence committed to use of LibreOffice and ODF, citing vendor lock-in [S-12].

However, around 2020, the stated motivations for switching begin to include explicit mention of digital sovereignty. The obvious proximate cause for this shift is the explicit top-down emphasis on digital sovereignty in the EU described in the February 2020 Commission and the July 2020 EPRS briefing mentioned in Section~\ref{sec:background-digital-sovereignty}. In 2020, the German state of Schleswig-Holstein released a report on open source software announcing a planned transition to open source, citing not only cost, but digital sovereignty [S-16]. In 2024 Schleswig-Holstein followed through with the announcement of the digitally sovereign IT workplace in state administration, involving switching to OSS office suites, operating systems, cloud, telephony and more software categories, citing not only the high costs of the de facto standards, but digital sovereignty, control over data, and investment in the local economy [S-18]. In 2021 Dortmund in Germany committed to using open source where possible, citing cost efficiency but also the principle of ``public money, public code'' [S-17]. In 2025, the Denmark Ministry of Digital Affairs announced a switch to LibreOffice, citing digital sovereignty, but also elaborating on how lock-in creates vulnerability and risks to freedom [S-21]. Similarly, in 2025 the French city of Lyon announced the ``Open Digital Territory'' and a coordinated use of OSS, citing digital sovereignty and environmental concerns [S-22].

In 2025, the solutions proposed also begin to broaden. The Dutch parliament passed 8 motions to support goverment switching away from de facto standards, citing digital sovereignty and local investment as motivations, as well as an apprehension of the contemporary administration of the USA [S-19]. Importantly, the solution proposed was not to use open source, but to invest in the private sector and nationalised home-grown alternatives. Similarly, in 2025 the municipalities of Copenhagen and Aarhus annouced the phase out of the de facto standards, citing digital sovereignty and uncertainty under the contemporary administration of the USA, switching to European, potentially private sector alternatives [S-20].

There are exceptions to this general timeline. For instance the Valencia region of Spain in 2013 completed a deployment of LibreOffice and LliureX (a Linux-based operating system) [S-9], citing costs, but also motivated by promoting the local ICT industry, the Valencian language, and promoting competition -- which anticipate the later concerns of digital sovereignty. Similarly in 2017, the City of Barcelona migrated towards Linux, OSS office alternatives, and coordinated use of OSS, citing costs but also promoting the local ICT industry [S-14]. The 2016 Digital Plan from the City of Barcelona outlines the plan for coordinated use of OSS, and while lock-in is cited as a primary motivation, the concept of ``technological sovereignty'' makes an early and explicit appearance here [S-13].

\paragraph{Changes in Motivation}
The general shift in motivation can be characterised as follows. Early switches were primarily motivated by practical concerns: the high costs of de facto standards and the avoidance of vendor ``lock-in''. However, later switches (particularly after 2020) were motivated more by ideological concerns: preservation of digital sovereignty, the avoidance of risk due to perceived instability or hostility of foreign administrations, the appropriate use of public funds, and investment in the local economy.

This might be characterised as a shift from instrumental rationality to value rationality, in Weber's terms \cite{weber1978economy}. Lower costs and freedom of dependence from specific vendors are clear expectations, and switching has a high likelihood of success in achieving these outcomes. Such switching actions can clearly be characterised as instrumentally rational (\emph{zweckrational}), that is, actions taken to pursue some specific ends. However, the ideals of stimulating the local economy, reduction of risk, and digital sovereignty and appropriate use of public funds in particular, are less amenable to objective measurement. While some aspects of these switching actions are still instrumental, they can be characterised to a much greater extent as being value rational (\emph{wertrational}), that is, actions that have intrinsic value for their ethical, aesthetic, religious, political -- in short, ideological -- symbolism. 

\paragraph{Changes in Solution}
A broadening of proposed solutions accompanies the shift in motivations. Early switching events focused heavily on adopting OSS standards such as LibreOffice and OpenOffice, and adopting the Open Document Format. While these still remain an important aspect of later switching events, the portfolio of solutions is notably expanded to include investments in software development initiatives that span the domestic public and private sectors.

Part of the reason for the appetite for de novo investment is a firsthand experience with the challenges of the open source model: its lack of centralised organisation, its fraught relationship with finance, its dependence on volunteerism, as elaborated by Eghbal \cite{eghbal2016roads}. Not all attempts to switch to OSS have been successful, as in the cases of reversion in the city of Munich [S-23], the Lower Saxony police and tax administration [S-24], and the city of Vienna [S-25]. Concomitantly we observe a shift in European government initiatives from coordinated use of free software to actively funded development including the private sector (discussed further in Appendix~\ref{apx:legislative-shift} and Section~\ref{sec:discussion-adaptation}).

\section{Study 2: Digital Patriotism in Consumer Software Adoption}
\label{sec:study2}

The analysis in Section~\ref{sec:study1} provides a brief evidentiary account of how geopolitically precipitated ideologies are surfacing in software adoption decisions in government agencies. We now turn our attention to whether, and how, such geopolitics may affect (and in turn be affected by) software adoption decisions on the part of individuals and businesses.

To study this, we conducted a qualitative analysis of discussions posted on the online forum Hacker News. Our methodology takes a ``small stories'' approach \cite{georgakopoulou2017small}, seeking to identify micro-narratives shared online about how commenters' own experiences of and motivations for software adoption have been affected by geopolitics and ideology. We take inspiration from prior work that has identified sites such as Hacker News as important sites for observable discourse \emph{in particular} regarding programming practice and software \cite{wu2014exploring, barik2015heart}. Similar analyses have previously been applied to understand the role of play in programming \cite{barik2017expressions}, spreadsheet features \cite{sarkar2022end}, and the changes in programming practice associated with the use of code-generating large language models \cite{sarkar2022programmingai, sarkar2025vibecoding}.

We note that the user base of Hacker News is historically skewed towards software industry professionals. We thus analyse these comments as one source of empirical signals about what technologically engaged Europeans value, while being cautious not to generalise to the entire population. This limitation is discussed further in Section~\ref{sec:discussion-limitations}.

This analysis was conducted following ethics guidance on internet data research from our institution. Hacker News comments are explicitly intended for public consumption. Under the platform's terms,\footnote{\url{https://www.ycombinator.com/legal}} users grant the platform a license to distribute this content, which the platform currently does through a public API,\footnote{\url{https://github.com/HackerNews/API}} which is provided to facilitate analysis of Hacker News data. In this paper, we pseudonymise participation by replacing usernames with numeric participant IDs, and redacting personally identifiable information from quotes. For reproducibility and provenance, we include a list of the thread URLs that were analysed. We acknowledge that it may be possible to recover username and redacted information by searching for text matches on the platform or in the linked threads. This is acceptable given the expectations of privacy and terms of use of this platform, and aligns with established research practices \cite{barik2015heart,barik2017expressions}.

\subsection{Method}
We retrieved an initial list of matching comment threads using platform-native search on Hacker News using variations of the keywords ``European alternatives'' and ``Euro stack''. Each thread was manually reviewed and included if the thread contained discussions pertaining to software adoption choices. This resulted in an initial dataset consisting of 1500\footnote{This is the exact number, not rounded.} comments from 762 unique authors (usernames) across 36 threads, collectively over 96,000 words. The thread URLs are listed in Appendix~\ref{apx:hn-threads}.

The dataset was qualitatively analysed in two phases. In the first phase, a highly directed content analysis \cite{hsieh2005three, assarroudi2018directed} was performed to annotate each comment according to the following binary categories: whether the comment was off-topic; whether the comment directly mentioned a personal anecdotal experience; whether the comment mentioned a motivation for seeking a software alternative; whether the comment mentioned challenges or limitations associated with seeking software alternatives; and whether the comment explicitly invoked the geopolitical context.

The initial analysis phase found that 737 comments (49.13\%) were off-topic, such as digressions about specific technologies, or debates around broader geopolitical issues that were not specifically connected to software adoption. After filtering these out, the final dataset consists of 763 comments from 514 unique authors (usernames) across 30 threads, collectively over 51,000 words. They spanned the date range 27 July 2011 to 16 April 2025. Of these, 112 comments directly mentioned personal anecdotal experiences, 229 mentioned motivations for seeking alternative software; 223 mentioned challenges or limitations associated with alternative software, and 146 explicitly invoked the geopolitical context for software adoption.

In the second phase, a thematic analysis \cite{braun2006using} was conducted of the entire final dataset, to identify motivations for seeking alternatives, challenges and limitations encountered, and aspects of the consumer consciousness of the geopolitical context for software adoption. Particular attention was paid to the comments reporting rationales for specific software choices made by the commenter, as these are (taken at face value) the most direct evidence of actual behaviour change being driven by (or at least associated with) geopolitical factors.

\textbf{Positionality}. The authors are based in the United Kingdom, holding affiliations at UK academic institutions and professional ties to a multinational technology company that is considered a de facto standard vendor in several of the software categories discussed in this paper. This positioning offers familiarity with the commercial logics of software provision, and a British perspective on European national identity and political dynamics. The authors prior work on postcolonial dimensions of AI data labour has cultivated a sensitivity to sovereignty discourses, which has attuned the analysis to the contested nature of digital sovereignty, while also guarding against uncritical endorsement of any single national or regional framing. We do not identify with a position for or against digital patriotism. Complete neutrality on questions of national identity and geopolitical alignment is neither achievable nor claimed; the interpretive choices made reflect an effort toward balanced analysis that readers should evaluate on its own terms.

\subsection{Results}

We find that commenters cite several motivations for seeking alternatives to de facto standards (Section~\ref{sec:results-motivations}). These begin with not meeting practical requirements, negative sentiment towards American technology companies, and apprehension of the American government. Going further, they include the desire to promote a diversity of values, to promote values considered virtuous, and nationalistic sentiment, that combine in a phenomenon we term \textbf{digital patriotism}. 

Commenters also encountered various challenges when seeking alternatives (Section~\ref{sec:alternatives-challenges}). They perceived the available alternatives as inferior, noting the complexity of replicating the de facto standards, perceived a lack of genuine alternatives, were skeptical of the benevolence of European companies, and were concerned about the capture of European identity by the European Union. 

Commenters speculated on the challenges of the European software landscape (Section~\ref{sec:results-europe-challenges}), such as its risk-averse culture, regulation and bureaucracy, lack of funding, fragmentation of languages, cultures and markets, hiring, and network effects.

Finally, commenters were conscious of the historic and geopolitical context of the software landscape (Section~\ref{sec:results-geopolitics}), invoking the post-World War II political order, differences in work culture and quality of life, and the ideological tradeoffs made by capitalist and socialist modes of organisation.

\subsubsection{Motivations for seeking alternatives}
\label{sec:results-motivations}

\begin{figure}[!tb]
  \centering
  \includegraphics[width=\linewidth]{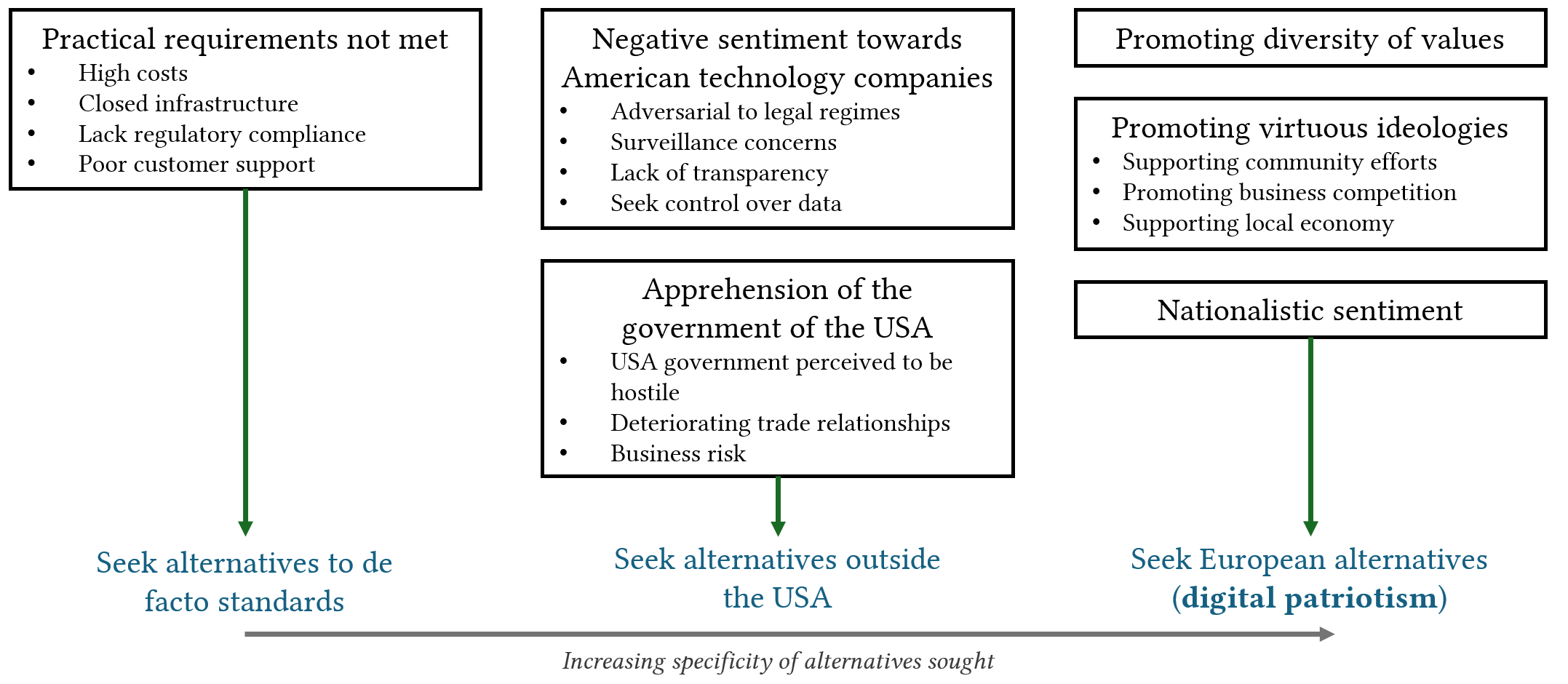}
  \caption{Motivations for seeking alternatives to de facto standards observed in our dataset, arranged from left to right in order of increasing specificity of alternatives sought. Explained in detail in Section~\ref{sec:results-motivations}.}
    \label{fig:alternatives-motivations}
  \Description{The figure is a structured diagram with three vertical columns: left, center, and right. Each column contains one or more rectangular text boxes that describe motivations. A green downward arrow in each column connects those motivations to a single action statement at the bottom of that column. Along the very bottom, a horizontal grey arrow runs from left to right with the caption “Increasing specificity of alternatives sought,” indicating that the actions become more specific from left to right. The boxes and text are in black, the arrows are green, and the bottom axis is grey.
    In the left column, a single box is titled “Practical requirements not met.” It explains that users encounter high costs, closed infrastructure, lack of regulatory compliance, and poor customer support. A green arrow leads down to the action label “Seek alternatives to de facto standards,” implying a broad, non‑specific search for options driven by unmet practical needs.
    The center column contains two stacked boxes. The upper box is titled “Negative sentiment towards American technology companies,” describing that such companies are seen as adversarial to legal regimes, raise surveillance concerns, lack transparency, and motivate users to seek greater control over their data. Beneath it, a second box titled “Apprehension of the government of the USA” states that the U.S. government is perceived as hostile, that trade relationships are deteriorating, and that this creates business risk. A green arrow from these motivations points to the action “Seek alternatives outside the USA,” reflecting a more targeted avoidance of U.S. options.
    The right column presents three boxes. At the top is “Promoting diversity of values.” Below it is “Promoting virtuous ideologies,” which emphasizes supporting community efforts, promoting business competition, and supporting the local economy. The third box reads “Nationalistic sentiment.” The green arrow in this column leads to the most specific action: “Seek European alternatives (digital patriotism),” indicating a preference for European options grounded in value alignment and nationalism.}
\end{figure}

Commenters cited a variety of overlapping motivations for seeking alternatives to the de facto standards. These motivations can be arranged in order of the increasing specificity of the requirements upon the solutions they correspond to, in other words, in order of how constrained the alternatives sought are. This is depicted in Figure~\ref{fig:alternatives-motivations}.

Commenters may be dissatisfied with the de facto standards for \textbf{not meeting practical requirements}. These include dissatisfaction with pricing models, that lead to perceptions that the de facto standards have high costs (e.g. \pquote{93}{I wanted to rotate a PDF 90 degrees \el and [de facto standard] wanted to charge for it, then I found that [European alternative] could do that and a whole lot more like edits and signature collections. I ended up subscribing \el the free tier already does several things that [de facto standard] would charge you for}). Closed infrastructure design, and contracts lead to perceptions that de facto standards lack flexibility and create vendor lock-in (e.g. \pquote{300}{For a startup, locking into an API that is initially free can often break their neck in the long run \el switching providers is a big distraction.}). The de facto standards may be non-compliant with local or EU regulations (e.g. \pquote{276}{We migrated to [European alternative] pretty easily and the main argument was that it's easier to conform to EU regulations if all your servers and data are on an European cloud.}). They might have poor customer support (e.g., \pquote{259}{One alternative that we are using is scalingo as a replacement for heroku. \el Their live chat is also an upgrade over support tickets.}). While these are important motivations to seek alternatives to the de facto standards, they do not necessarily entail a shift to European alternatives, or even away from American software. It may well be possible to find American alternatives that are less expensive, with more flexible contracts or with greater interoperability, with better regulatory compliance, and better customer support.

The motivation to move away from American vendors in the search for alternatives derives from two interrelated sources of dissatisfaction: a generalised negative sentiment towards American technology companies, and a generalised apprehension of the government of the USA.

Commenters described a generalised \textbf{negative sentiment towards American technology companies} (e.g., \pquote{397}{The startup I'm currently working for doesn't even consider US based tech solutions at the moment.}). Commenters indicated that they perceived the vendors of the de facto standards as being adversarial to legal regimes (e.g., \pquote{340}{US tech industry \el work hard on malicious compliance.}). Further motivations that contributed to this sentiment included concerns over surveillance (e.g., \pquote{164}{if you think your internet traffic isn't being tapped by the US regardless you're completely delusional}), privacy (e.g., \pquote{253}{I found [... European providers that have] Significantly more convincing privacy story than US providers}; \pquote{431}{My employer moved away from American-hosted products due to privacy concerns}), and lack of transparency (e.g., \pquote{12}{maybe two or three times a month, paypal initiates an "investigation"  which usually results in the transaction being reversed \el There is really no transparency}). Consequently, commenters sought alternatives that gave them greater control over their data (e.g., \pquote{398}{I [... switched to a European alternative because] I'm not going to keep putting all my notes and personal data in the hands of American companies}).

Similarly, commenters described a generalised \textbf{apprehension of the government of the USA}. 
Several commenters perceived the US government to be hostile (e.g., \pquote{253}{I [... switched to a European company] as the US authorities start to look increasingly unpredictable)}; \pquote{398}{I [... moved my data to a European company because] American companies \el might be pressured by the fascist US govt to hand them over}). 
Commenters cited the deteriorating trade relationships between the USA and other countries as motivation to seek alternatives (e.g., \pquote{263}{Having previously run a >200 person software company \el I sure would have a migration plan ready and alternative explored \el it's entirely plausible that US services are either hit with things like tariffs, or end up suffering US state censorship}; \pquote{434}{In a trade war it is possible that digital services are hit with tariffs [... so European alternatives are sought because] This will not happen with EU products.}). 
This contributes to a heightened sense of unpredictability and business risk due to exposure to American technology and by extension the US government (e.g., \pquote{262}{you're going to want to mitigate the risk of US services now that the US is an adversary \el It wouldn't be far fetched to imagine the EU making it illegal to use US tech companies similar to how we can't use Chinese or Russian tech companies.}; \pquote{315}{having dependencies on services provided by the US [... has become] a serious risk}; \pquote{375}{It's not just a consumer movement, it's Europe itself moving to get away from an unreliable schizophrenic ally. You can't make long-term plans when every 4 years you're playing russian (eh) roulette with your partner.}).

Thus, commenters with motivations of the above two types seek software from outside the USA. However, this still does not necessarily entail that the solution must be a switch to a \emph{European} alternative. There are other continents and countries with thriving software industries that may be looked upon as sources for alternatives. Open source software may also be a potential solution. To make the final step towards the search for specifically European alternatives, commenters cited a variety of ideological motivations that come together in a phenomenon we term \textbf{``digital patriotism''} (explained shortly). 

Several commenters emphasised the importance of seeking alternatives as \textbf{promoting a diversity of values} (e.g., \pquote{263}{I recently moved my software business mostly off US services. Diversifying away from the US is one of the most important things all of us outside the US can do for the free world right now.}). Another commenter describes their choice of social network based on the content moderation policy: \pquote{248}{As long as [the service] doesn’t start censoring topics which the [US] government now considers controversial}.

Going further, commenters emphasised the importance of \textbf{promoting values and ideologies that they considered virtuous} (e.g., supporting the open source community: \pquote{462}{I've started using [... a European alternative] - great company, great service, they've recently donated a [... large] server to the Debian project as well (including hosting)}; \pquote{465}{I've been using [a European company] for a decade \el they actively work to support the Open Source community}). 
Another example of a virtuous ideology that was the basis for seeking alternatives was promoting business competition (or resisting anti-competitive behaviours) (e.g., \pquote{156}{rich companies can just bribe their way into people's computers by offering their games for free. \el I won't install that just on principle alone. [... I] chose another game and I wish many more will do that}; \pquote{118}{I prefer supporting competition in the market, because I know that as soon as monopoly gets established it starts abusing all marlet participants and acting like a little dictator.}).

As another common example, several commenters emphasised the importance of supporting local businesses and the local economy (e.g., \pquote{433}{If a local (national) product is comparable to a foreign one, I’ll usually choose the local one.}), often making the argument that spill-over effects from supporting the indigenous software sector would support the local economy as a whole (e.g., \pquote{119}{Local business means local taxes. So by supporting local businesses near your place you might indirectly profit from better infrastructure, better school etc. (things which are paid by the government with these local taxes).}; \pquote{124}{If you buy from a smaller local euripean [sic] provider they will [pay] local taxes. Local taxes pays for schools, roads and nurses.}).

Finally commenters cited \textbf{nationalistic sentiment} in motivating the transition away from de facto standards (e.g., \pquote{514}{We are based in Germany \el it feels good to be here and have our server also hosted here with a German company}; \pquote{83}{I find the UK's inability to compete with the US web giants depressing. I don't like the idea of relying on a few American companies \el I'd love to see us build a British Google for example.}; \pquote{339}{this [search for alternatives] is the best thing that could possibly happen for Europe long term. Shed the dependency on the US \el this US adminstration is the single best thing that has happened to Europe and the UK. This is where Europe gets reborn.}).

It is this final set of motivations that leads specifically to the search for European alternatives (and indeed, in some cases, to search for alternatives from specific European countries, although our evidence for this is limited -- we acknowledge this briefly in Section~\ref{sec:discussion-limitations}). Digital patriotism strikes us as a novel and important phenomenon. As an ideology, it is the counterpart at the individual and business level to the ideology of digital sovereignty at the level of the State. Indeed, some commenters explicitly identified that the discussion of software adoption choices is renewed by Europe's push for digital sovereignty (e.g., \pquote{348}{Talks about investing 800 billion in the military industry across the EU would have been unthinkable 4 months ago. Now they're proceeding at a speed that is staggering \el Holland for instance voted and accepted a resolution last week for more digital sovereignty. This doesn't mean we'll have an [de facto standard] competitor overnight, but we're also only 3 months in.}).

\paragraph{Digital patriotism} Our choice of the word ``patriotism'' makes a deliberate contrast to the deeply related concept of nationalism. Patriotism differs from nationalism in a two aspects that we wish to centre with this metaphor. First, patriotism acknowledges the weaknesses of one's State and is grounded in a desire for its betterment, whereas nationalism is associated with a belief in the inherent superiority of one's State \cite{anderson2020imagined,gellner2015nations,kosterman1989toward,scruton2007palgrave}. Second, perhaps consequently, patriotism entails making sacrifices \cite{kosterman1989toward,scruton2007palgrave,berns2002making,samsu2022understanding}.

We observe both these distinctions in our dataset. Not only were commenters acutely aware of and critical of the limitations of European alternatives to de facto standards (detailed in Section~\ref{sec:alternatives-challenges}), but they were willing to make compromises in their choice of software, i.e., sacrifices, to serve ideologies that transcended instrumental values (e.g., \pquote{84}{[European alternative] don't provide a webmail as good as [de facto standard], but they provide great email service while focusing on user privacy.}; \pquote{221}{At my (enterprise) job there has been a noticable shift in attitude. We've just decided to use [European alternative] for OCR, even though everything we currently use is embedded in [de facto standards]. Usually any push for EU products would be dismissed because of the initial friction compared to SV products. This seems to have shifted significantly all the way to the top. Everyone seems on board to just accept more friction to use EU products. We are now even moving to a [European cloud alternative]. This was unheard of even a few months ago.}).

We acknowledge that the concept of digital patriotism is at an early, exploratory stage; it overlaps with localism in ways that are difficult to disentangle given our dataset, so we deliberately avoid claims of predictive power. We treat digital patriotism as a sensitising concept \cite{blumer1954what} or ideal type \cite{swedberg2018use} that orients analysis without prematurely fixing attributes or measures. Constructing and validating quantitative survey instruments for measuring digital patriotism would be premature and potentially distortive at this stage. However, such operationalisation is suitable for future work.

\subsubsection{Challenges and limitations encountered when seeking alternatives}
\label{sec:alternatives-challenges}

As evidenced in the previous section, many commenters shared firsthand motivations for searching for alternatives to the de facto standards. However, this process was not always straightforward. Many commenters shared challenges and limitations they encountered when switching.

The primary category of challenges was associated with the quality of the software alternatives themselves. Several commenters \textbf{perceived the available alternatives as inferior} (e.g., \pquote{467}{[The European cloud alternative we use] is good, the hardware is (mostly) good, the provisioning is fast, the support responses are fast [... but] their network is terrible}). 
Most commonly, alternatives were seen as lacking feature parity with de facto standards (e.g., \pquote{27}{After having switched [... to a European email provider] \el I only wish they supported real 2FA.}; \pquote{259}{One alternative that we are using \el doesn't have all of the same features}; \pquote{450}{I never found [... a certain European alternative] customisable - I mean it doesn't even work on third party apps.}).
They could also be more expensive (e.g., \pquote{473}{I have found the experience of using [European alternative] quite good, but it is much more expensive than other services.}; \pquote{508}{[European alternative] are good. But still dont match [... the de facto standards] in terms of pricing.}; \pquote{488}{The only downside [of a certain European alternative] is: Prices are currently still higher}).
Some commenters who had attempted to switch to open-source alternatives noted the poverty of open-source solutions, mirroring the development of governmental attitudes towards open-source as observed in Section~\ref{sec:study1} (e.g., \pquote{94}{I am continually frustrated at just how difficult of a problem simple PDF operations are \el The FOSS tools aren't great}).

Commenters, particularly those with apparent experience in the software industry, noted the \textbf{huge complexity of replicating the features of de facto standards} (e.g., \pquote{22}{If someone offered European customers better service than Paypal at roughly the same price, I'm sure it would be popular. But it turns out that handling transactions in 30+ countries is pretty complicated and fraught with bureaucratic problems, which is why Paypal remains the only option.}; \pquote{275}{[European alternatives] often cannot compete on price or on capability. The scale and mature feature sets of US based cloud providers outweigh the benefits of keeping it within European ownership for most commercial organizations.}; \pquote{86}{Here in France we have [a search engine] \el that never really worked \el they were just providing a proxy for Bing's results \el it was people with money that said "we are going to be the French Google". But \el being "the French Google" is not enough.}; \pquote{173}{Mojeek is a startup vs the trillion dollar valuations of Bing and Google [... who also possess a] vast history of query data that no one else has.}).

This built up into a general perception that there is a \textbf{lack of genuine alternatives} to de facto standards (e.g., \pquote{416}{Why do the European alternatives suck so much? - Pretty much all cloud providers seem (very) subpar to AWS. \el There's no real Cloudflare alternative. - There's no real Google Workspace / Microsoft Teams competitor (Mail + Calendar + Drive). - No search engine / Google alternative.}; \pquote{20}{people still use Paypal even after all the horror stories: there is no alternative.}; \pquote{263}{In looking for non-US alternatives, I [... found] a lack of solid github/gitlab competitors.}), and European software was perceived as uncompetitive (e.g., \pquote{117}{why not come up with products that can compete in the global market? \el most EU tech is just garbage.}; \pquote{164}{[A leading European email provider] is \el not [a] patch on Gmail,O365, iCloud, Zoho or one of the plethora of other services that hail from the US.}).

Another set of concerns centred around the \textbf{strategic responses made by the de facto vendors} in complicating the decision to seek alternatives. These included mitigations such as efforts of the de facto standards to comply with European regulations (e.g., \pquote{109}{what does European mean? \el Does that make you European \el if you are say a non-European company but meet all the European data protection and compliance requirements and the European counterparts do not?}), or efforts to move some operations to Europe (e.g., \pquote{431}{My employer moved away from American-hosted products due to privacy concerns, except for [... one de facto standard, where] we have the standard assurance \el that our data is in Ireland}). By reducing the perceived business risk of the de facto standards, such responses make it more difficult to convince decision makers within organisations of the marginal benefits of seeking European alternatives. We discuss some of the adaptations made by de facto vendors in greater detail in Section~\ref{sec:discussion-adaptation}.

A final, major set of concerns was \textbf{skepticism about the benevolence of European companies} (e.g., \pquote{188}{[An email provider] were supposed to be the go to private email back in the day. Turns out the FBI was running the servers for who knows how long. I’ve had similar concerns about [European alternative], looks like I’m not the only one.}; \pquote{451}{[European alternative] provided user information in 10,368 requests in 2024}). 
Indeed, commenters noted that European companies, beyond the requirement of participating more fully in the European regulatory regime, had no \emph{ipso facto} commitments to other values and ideals that motivated the search for alternatives, values such as promoting competition and prioritising consumer rights (e.g., \pquote{185}{Avoid to sympathize for companies based on non-competitive parameters, they aren't your friends or do anything for you regardless of where they are}; \pquote{322}{This is just shifting your eggs from one basket into another. Yes, the current basket is on fire but we should assume any basket can be lit on fire just as easily.}; \pquote{457}{Yeah, try these [alternatives] if you prefer EU surveillance. So far, the anti-E2EE proposals have been rejected, but we don't know for how long that will last.}; \pquote{511}{The UK is in the privacy-conscious EU and apparently \el taps all the internet traffic going in or out of Europe.}; \pquote{492}{I would also be wary about having servers on some EU countries and check on their relationship with USA in data sharing.}; \pquote{460}{it [is] not about fostering a dynamic European market \el but it sounds more like, Hey, let's use the services from our own European cartel companies and help cement them}).

Related to this was the concern of the idea of ``Europe'', and the European Union in particular, \textbf{capturing European identity} and subsuming the identities of its constituent nations (e.g., \pquote{49}{It's quite off-puting branding, playing with the flag of a bureaucratic institution like the EU. If it was about what Europeans create, I would be receptive. Instead, they are promoting whatever is in the European Commission agenda, which is radically different to what I need.}; \pquote{261}{despite the European Commission's attempt to claim the term Europe for itself, the EU and Europe are not the same thing}; \pquote{433}{I feel exactly the same kind of “connection” to Europe as I do to the USA. It’s a different story when it comes to the country I live in. \el But all this EU vs USA fuss means nothing to me.}), which complicates the moral positioning of EU-based alternatives.

\subsubsection{Perceived Challenges of the European Software Landscape}
\label{sec:results-europe-challenges}
Additionally, in response to the perceived lack of genuine alternatives as outlined in Section~\ref{sec:alternatives-challenges}, commenters also speculated in depth about the possible reasons for this, and the broader challenges faced by the European software landscape.

Several participants perceived Europe as having a risk-averse culture (e.g., \pquote{34}{The US, especially Silicon Valley, has a culture of taking risks, of venture capital \el Europe tends to be more risk-averse. We want to invest, but only once it's clear it's going to be a success. People are much less eager to take risks with their money, and entrepreneurs are less eager to risk bankruptcy.}; \pquote{78}{The biggest problem is funding due to the risk aversion of EU institutions and limited alternative sources. \el You can't fail in the EU because you only have one shot and won't get back again with another company/idea. In reality, it's very likely you'll stumble a lot initially and won't have the leeway that you have in the US.}), and a culture of business conservatism that disincentivised competition (e.g., \pquote{49}{[European companies are] in no rush to become more efficient. \el Compare this with the cut throat competition in Sillicon Valley \el That's how you end up with non existent tech offerings in Europe.}; \pquote{350}{delivering on the cut-throat tech competition on a reasonable time frame seems not part of the european ethos.}).

Others noted challenges in regulation and bureaucracy (e.g., \pquote{44}{I find the amount of legal red tape you need to deal with incredibly discouraging. \el I have the capacity and funds to start a business, but I just don't think it's worth the hassle.}; \pquote{418}{We’re building [... a European alternative] \el Being EU based and trying to target only EU customers is more complicated, from fundraising to customer acquisition, especially when you factor in language localization and legal bureaucracy.}), noted the high costs and low ease of doing business (e.g., \pquote{77}{[There are issues with] Cost of incorporating; cost of hiring and firing}; \pquote{57}{[European] countries slap so much red tape and so many fees on this.}; \pquote{362}{Endless stupid regulations and crippling taxation make startups impossible.}), ultimately attributing this to the expanded functions of the State (e.g., \pquote{48}{Big Government is probably a big part of it \el The public sector is so big that many software engineers can spend their entire career as overpaid government contractors.}; \pquote{59}{In [... Europe] running a business is harder because there's an expectation that businesses can take care of their workers. A lot of red tape exists to make sure that a business can shoulder these things. \el There really are a lot of implications to creating the sort of environment that the US or to a lesser extent the UK has, and it doesn't work well with the economic model of most of Europe.}).

Others noted a lack of funding (e.g., \pquote{399}{In my experience \el you can grow huge in the USA market, that is where the funding goes}; \pquote{418}{We’re building [... a European alternative] When you’re fundraising, you’re always compared to global competitors, not just European ones. Which usually have better access to capital}; \pquote{76}{It's capital. We simply don't have it [in Europe], or it's not allocated to tech.}), contrasting the situation in particular with the highly developed venture capital market in the USA (e.g., \pquote{42}{The US has a very mature and developed tech VC scene. Where's Europes isn't as mature.}; \pquote{76}{Europe has no VC.};).

Others noted that various kinds of fragmentation across Europe may increase the challenge for the European software landscape to incubate leading alternatives to the de facto standards. This included the wide variety of European languages and cultures (e.g., \pquote{80}{Language is a big barrier. \el we had to build a lot of custom tools to narrow the linguistic gap.}; \pquote{77}{fragmented market in terms of policy, culture and language}), the small and fragmented individual markets (e.g., \pquote{30}{the biggest problem is the small home market / fragmented market. While the EU has a very well integrated market for industry goods, for services (where I would include software) the market is less than perfect.}), and fragmented regulations (e.g., \pquote{36}{In the US \el contracts are easy to read because they're mostly standard. In the EU, there's a lot of [... regulatory heterogeneity] By the letter of the law, in the worst case, our business in the EU could not have existed. \el The standardization in the US makes it easy. But if I ever have to figure out something for Germany again, I'll slit my throat and feed lizards with my blood first.}; \pquote{33}{common law/company to easily hire people across EU would help.}).

Others reflected on the challenges faced in hiring (e.g., \pquote{38}{Hiring SE in Sweden takes minimum 6 months. Take into account a Union to deal with. It's probably too complex for some startups.};), noting stark differences in compensation structures (e.g., \pquote{69}{software engineers don't get paid nearly as much as in the US.}; \pquote{194}{EU companies are willing to pay for top talent too, it's just that the amounts they can afford to pay are smaller than US companies}).

Finally, many commenters noted that the de facto standards have first-mover advantages, which produces network effects to drive adoption (e.g., \pquote{21}{alternatives cannot gain ground because everybody uses paypal. It's a vicious circle.}; \pquote{388}{User stickiness is hard to break due to providers' walled gardens, even for me who is aware and willing to change.}; \pquote{127}{networks have no replacement. At the end of the day, social sites live and die by the circles of people that are there.}). The de facto standards also enjoy customer familiarity (e.g., \pquote{61}{Even when there are local apps available, many small businesses (and larger ones) will stick with services from big, well-known tech companies \el because of inertia, or simply because those products feel safe and familiar.}) and in some cases are viewed as having superior marketing (e.g., \pquote{119}{US companies are much, much better at marketing their solutions. Often even the european media landscape is only mentioning US services in  reviews and comparison.}). This creates a vicious circle for European alternatives, who lack a similar track record.

\subsubsection{Consumer consciousness of historic-geopolitical context}
\label{sec:results-geopolitics}
Commenters showed a strong consciousness of the historic and geopolitical context within which their software adoption decisions are made. In particular, as Hacker News commenters are biased to being software industry professionals, discussions of the challenges of European alternatives and the perceived scarcity thereof often naturally led to participants debating rationales for preferring to live and work in Europe and thus contribute to growing its software industry.

Several commenters raised aspects of the historical relationship between Europe and the USA to explain the status quo of the software landscape. In particular, many commenters traced the economic and ideological divisions to the aftermath of the second World War (e.g., \pquote{42}{Europe was simply a mess in the aftermath of WWII there weren't going to be many tech revolutions taking place there \el there was such a congregation of talent in Berlin in the 1930s that some have predicted a second "Renaissance" was inevitable were it not for WWII.}; \pquote{54}{The US had way more wealth compared to Europe, thanks to unbridled capitalism in the past centuries and not hosting a World War. That wealth trickle down and converted in various shapes until the current VC class.}; \pquote{165}{Marshall Plan essentially fueled demand for american goods, because funds provided in US Dollars pretty much translated to purchasing things that could be sold for USD. The network effects were huge and paved the way for dot com boom and later cycles}; \pquote{332}{ Europe is a de facto vassal state to US since WW2. What Europe didn’t understand or didn’t want to understand was the US agenda differs from Europe agenda , always has}).

The landscape of the software industry and decisions around software adoption were seen, not as distinct and isolated processes, but historically and geopolitically contingent, and connected with differences in work culture and quality of life, and the broader debate between capitalist and socialist modes of socio-economic organisation (e.g., \pquote{83}{I'll grant that the US is ahead in big tech but I'd never move there. The quality of life is pretty poor compared to wealthy European countries.}; \pquote{90}{I don't need to worry about my family in Europe. No matter what health problems they have, they will be taken care of in a good public hospital and it won't cost them a thing. When my kids are old enough to go to university, they won't have to take out a loan to pay for courses. If they drop out of college, they'll still be able to get a decent job that pays a living wage. They'll have access to affordable child care for their children, and they won't need a 200.000€ job just so they can afford living close to their job. I don't have to be afraid of getting mugged or shot on the street if I go to the wrong part of town because we have much lower crime rates because  our social security system helps people who run into hardship instead of punishing them.}; \pquote{437}{excessive centralized control, even in a market system, can stifle economic dynamism \el the US consistently produces globally dominant firms, a sign of higher productivity and innovation. \el in large part because it avoids the EU’s regulatory morass and therefore has a more market-based economy.}). Thus, digital patriotism, as expressed through (relatively) minor acts of software adoption, is seen as commenters as being bound to broader expressions of identity and values that are expressed and enacted by much more consequential acts and choices of where to live and work.

With regard to the preceding observations in Sections~\ref{sec:results-europe-challenges} and~\ref{sec:results-geopolitics}, it is worth emphasising that these are perceptions, and these quotes should not be taken as a form of third-hand evidence for the reality of the European situation. Often these ``folk theories'' \cite{johnson1998basic,devito2018people} about the difficulties of Europe, such as the fact that it has ``small markets'' in comparison to the USA, can be refuted directly with counterexamples such as Israel (e.g., \pquote{40}{[The size of European markets] doesn't explain why Israel doesn't have this problem, they have an even smaller home market. They just sell internationally.}). Folk theories based on perceptions such as these may be refuted by the personal experiences of others, and may fail to convince them. For instance, commenters' perceptions of how ``bureaucratic'' Europe is versus America differed based on personal experience (e.g., \pquote{45}{I have a hard time to see how Europe would be any more bureaucratic then the US. Working for a European company that do business with American companies \el the contracts lengths and the legal staffing is a factor of ten higher every time we try to deal with an American customer. \el involving lawyers just to sign standard contracts is not even something we have to do with European customers}). This is also possibly because concepts such as ``red tape'', ``regulations'', ``bureaucracy'', etc., are extremely nebulous and may be experienced at so many different levels, and individuals may be directly exposed to vastly different regulations that may be experienced as more or less burdensome depending on their own context. But to the extent that these are real perceptions that influence software adoption, they create their own reality -- a hyperreality \cite{baudrillard1983simulacra}, if you will, that is more consequential to our concerns.

\section{Discussion}
\label{sec:discussion}

\subsection{From Digital Sovereignty to Digital Patriotism}
\label{sec:discussion-catalogues}
We have speculated a parallel between the evolution of motivations for switching away from de facto standards at the government level as documented in our first study (Section~\ref{sec:study1}), and at the consumer and business level as documented in our second (Section~\ref{sec:study2}). Namely, we posit that the ideology of digital patriotism is the individual correlate to the national ideology of digital sovereignty. However, the precise nature of the relationship between these phenomena remains to be explored. Does one, in any meaningful sense, follow from the other? Does the need for sovereignty ``trickle down'' in the form of patriotism to individual consumers and businesses, or does sovereignty perhaps emerge as the political consolidation of individual patriotic will refracted ``upwards'' through democratic process?

While this question is out of scope, we will note here some phenomena that appear to bridge the two. These are the network of community efforts to promote such ideologies. They include community fora such as the r/BuyFromEU subreddit \cite{reddit-buyfromeu} and the Buy European feddit \cite{feddit-buyeuropean}. They include end-user tools such as the Go European browser extension that aims to provide \textit{``European alternatives to non European websites''} \cite{k-robin-goeuropean}. They include efforts to catalogue alternatives to de facto standards ranging from simple lists and pages (e.g., \cite{dataethics2025european, privacyguides2025privacy}), to sponsored and diligently maintained databases (e.g., \cite{datenpunks-switching, bankrupt-alternatives, prismbreak} and notably \cite{graf-european}), to curated selections of companies, alliances, and networks such as Eurostack \cite{bria2025eurostack} (which has incurred critique of its own \cite{fioretti2025eurostack}). They also include open letters from the private sector such as \textit{``Stand Tall Europe''} \cite{rogaczewski2025stand}, and influential campaigns such as the Free Software Foundation Europe's \textit{``Public Money, Public Code''} \cite{fsfe-public}. A more detailed investigation of the role that such fora, tools, lists, databases, letters, and campaigns play in software adoption is suitable for future work.

\subsection{Adaptation and Co-operation with De Facto Standards}
\label{sec:discussion-adaptation}

The de facto standards, particularly those most exposed to regulatory and compliance requirements in Europe, are moving in response. 
For example, Oracle launched its EU Sovereign Cloud in 2023 \cite{rebmann2025_oracle_eu_sovereign_cloud}. 
Amazon publishes transparency reports disclosing whether they have been required to share data stored outside the U.S.A. with the U.S. government under the CLOUD act since 2020 \cite{aws_cloud_act}, announcing a digital sovereignty pledge in 2022 \cite{garman2022sovereignty_pledge}, following up in 2023 with the launch of the AWS European Sovereign cloud with infrastructure located wholly within the EU \cite{garman_peterson2023_eu_sovereign_cloud}, announcing a 7.8 billion Euro investment in the program in 2024 \cite{aboutamazon2024_eu_sovereign_cloud_investment}, and an independent advisory board for it in 2025 \cite{kunert2025_aws_euro_cloud}. 
In 2019 Google announced a commitment to publish CLOUD act requests \cite{googlecloud_advancing_customer_control}, and reinforced their commitment to investment in cloud infrastructure in Europe \cite{googlecloud_deepening_commitment_europe}, in 2020 outlined \textit{``three pillars of sovereignty''} in Google Cloud \cite{googlecloud_2020_data_sovereignty_europe}, in 2021 announced the \textit{``Cloud. On Europe's Terms''} program which partners with local technology providers to independently manage sovereignty controls and measures \cite{googlecloud_digital_future_europes_terms}, and in 2025 updated sovereign cloud services to include a fully standalone ``air-gapped'' platform \cite{theregister_2025_google_sovereign_cloud_updates}. 
In 2022 Microsoft announced a set of European Cloud principles and new initiatives to support European cloud providers \cite{smith2022_eu_cloud_feedback}, and the Microsoft Cloud for Sovereignty later that year \cite{sanders2022_cloud_for_sovereignty}. In 2024 the Windows operating system was updated to comply with requirements from the Digital Markets Act in the European Economic Area \cite{windowsinsider2023_dma_preview}. In 2025 Microsoft announced the completion of the EU Data Boundary for storing and processing data in the EU and European Free Trade Association (EFTA) regions \cite{brill_lorimer_2025_eu_data_boundary}, and later that year announced European digital commitments to invest in datacenters in Europe and a ``Digital Resilience Commitment'' to ensure the continuous operation of the same \cite{smith2025_european_digital_commitments}. 

In addition to the efforts to stimulate private sector innovation mentioned in Appendix~\ref{apx:legislative-shift}, which are largely aimed at domestic, often de novo enterprise, there appears to be a current of appetite for co-operation and collaboration with de facto standards. For example, in February 2025 the UK government (Department of Science, Innovation and Technology) signed a memorandum of understanding with Anthropic, developers of the Claude family of language models, to collaborate on research for supporting innovation and mitigating risks \cite{dsit2025anthropicmou}. A similar MOU was signed with OpenAI, developers of ChatGPT, in July 2025 \cite{dsit2025openai_mou}.

A related but distinct phenomenon is the co-opting of the language of sovereignty into the motivations of other private and third sector initiatives, beyond those of the de facto standards. For example, the non-profit trade association CISPE (Cloud Infrastructure Providers in Europe) announced in 2025 a 1 million EUR commitment to funding OSS development, endorsing \textit{``digital sovereignty principles''} such as being \textit{``independent from non-European government influence''} \cite{cispe2025_fulcrum_investment}. CISPE's activities include the Gaia-X program, which runs an endorsement system for certifying cloud solutions as offering \textit{``sovereign control over data''}  \cite{gaiax_about}. Similarly, a partnership in 2024 between German AI company Aleph Alpha and Finnish company Silo AI was described as being motivated by the shared \textit{``values on sovereign AI''} \cite{macaulay2024_alephalpha_siloai}, and in 2025 British company NScale and Norwegian company Aker in partnership with OpenAI announced Stargate Norway, a Norwegian AI gigafactory aimed at delivering \textit{``sovereign, scalable and sustainable infrastructure''} \cite{nscale2025_stargate_norway}.

Of course, such developments are not without skepticism. As Parker and Carter note, it is imperative that deeper partnerships between the private and public sector fulfil their mandates, and that \textit{``existing regulations are adequate to ensure that public sector AI works for everyone, supports public good and prioritises people''} \cite{parker2025licence}. Similarly, Grohmann and Barbosa observe that efforts to provide ``sovereignty-as-a-service'' may \textit{``frame sovereignty as a technical, legal, and infrastructural matter''}, transforming it into something that can be provisioned rather than exercised \cite{grohmann2025sovereignty}.

These currents, both the efforts of the de facto standards to ensure that their products can be viewed as consistent with the ideology of sovereignty, and the appetite for government collaboration, are noted here for their potential to complicate digital patriotism as an emerging rationale for software adoption. They are beyond the scope of this paper to explore in depth. Future work may investigate whether, how, and to what extent such efforts may affect the adoption or non-adoption of de facto standards as a mode of enaction for digital patriotism, or the role of government policy in either setting, or reflecting, public sentiment.

\subsection{Implications and Non-Implications}

The most salient implication is for the evolution of software adoption models to incorporate some notion of ideology or value rationality, and how precisely to do this is suitable for future work. The investigation in this paper has outlined one concrete example, namely, the ideology of digital patriotism that is in reciprocal configuration with the ideology of digital sovereignty, which in turn is a product of geopolitical currents. However, it is far from the only example. For instance, recent research has documented the use or non-use of Generative AI tools on ideological grounds \cite{sarkar2025aishaming}. We have also learnt that the use of software can be deeply bound up with personal and professional identity \cite{nouwens2018application, sarkar2022end, sarkar2023simplicity}.

As such, we observe that much software use that might formerly have been viewed as discretionary and instrumental is increasingly coming to be \emph{value-laden}. This marks an evolution, perhaps a signal of maturity, of the place of software in society: where the choice of technology can be appropriated as a symbolic gesture in the manner of the printing press in the Protestant reformation, the calendar in the French revolution, or Gandhi's \emph{charkha} spinning wheel in India's independence movement. It is a mark of maturity, because like the press, the calendar, and the \emph{charkha}, any technology must first be sufficiently assimilated into society so that its direct, first-order sign can be established, and only then can it acquire second-order ``mythic'' signification \cite{barthes1972mythologies}. We contend that such significations, now more acutely than ever before, attach themselves to software choice in the workplace.

A study like this, which characterises choices and actions made under political considerations, naturally begets questions about appropriate modes of political action. For instance, given our main conclusion -- that software adoption is entering a phase strongly influenced by geopolitical ideology -- we might wish to draw implications for questions such as ``How ought American businesses to respond?'', ``How ought European businesses to respond?'', ``How can we help consumers make more value-aligned decisions?'', ``How should policymakers and regulators take this into consideration?'' We deliberately avoid questions such as these within the scope of this paper. Speculating upon them requires certain value judgements around what positive outcomes look like, and which parties in apparently zero-sum scenarios one is vested in.

\subsection{Limitations}
\label{sec:discussion-limitations}
While we made extensive effort to search for suitable sources documenting switching events for the first study, it is possible that some sources might have been overlooked, particularly those that are only indexed by specialised search engines (e.g., on various municipal government portals) or which have no accompanying English documentation which might have enabled the English-speaking authors of this paper to discover it. Future work could adopt more systematic or diverse review methods.

The analysis treats stated motivations in public-facing documents as relatively transparent indicators of underlying rationales. However, government procurement rhetoric is strategic, performative, and often post hoc. Long-standing cost and lock-in concerns may have been superficially rebranded in the language of sovereignty after 2020, without a corresponding change in underlying decision-making logic.

Analyses of online communities are subject to self-selection bias; participation in our Hacker News is skewed towards software developers, who are a type of software consumer that take an intrinsic interest in the opinionated use of software. This group may not be representative of software consumers more broadly -- many of the software choices commonly discussed in our dataset centre around cloud hosting, payment processing, etc. which are decisions primarily made by developers and businesses. Future work may expand upon this by analysing sources with different representational biases (e.g., Reddit), or conducting surveys where the sample can be better controlled.

Nevertheless, precisely because Hacker News concentrates ``lead users'' \cite{von1986lead} and early adopters, its discourse can foreshadow broader preferences and trade‑offs that diffuse later to mainstream consumers \cite{kraishan2025launch}. Per‑capita participation snapshots suggest notable Hacker News engagement from multiple European countries \cite{hn2023_demographics_by_country}, making it a plausible (albeit still a professional, self‑selected) lens on Europe-centred technology debates. User‑generated discourse systematically shapes and reveals consumer preferences \cite{schroder2025unraveling}; thus, even a specialised community can surface emerging value priorities that merit follow‑up with complementary methods.

Finally, there are several limitations of scope; phenomena that are clearly in need of deeper treatment but the inclusion of which would be an infeasible expansion of the subject suitable for a conference paper. However, each of these matters would make suitable future work, particularly as it pertains to the study of ideological motives in software adoption:
\begin{itemize}
    \item Whether and why government switches succeed. In our first study we noted several switching events, but not all are successful in the long term, as can be observed in the reversions of the city of Munich, the Lower Saxony police and tax administration, and the city of Vienna (sources S-23, S-24, and S-25 respectively in Appendix~\ref{apx:government-switches}).
    \item Switches outside of Europe. We scoped the first study to switches observed in European governments. We note that such efforts are not limited to Europe -- for example, we encountered instances of switching in India \cite{rudra2023_indian_defense_linux, fpexplainers2023_maya_os, ec2025_eu_india_ttc_statement}, Canada \cite{paul2009_canadian_open_source_feedback}, and China \cite{kaur2022_kylin_china}. Consumers in these regions may make software adoption decisions based on more site-specific ideologies.
    \item Hardware. The manufacture and circulation of semiconductors has its own distinct geopolitics, which we were not able to connect to software adoption choices from the data used in our studies. 
    \item The precise line of what software adopters perceive as sufficiently ``European'' -- in our data for the second study we came across commenters raising questions such as whether software can still be considered European if it is acquired by an American company, or whether a search engine can be considered European if it is built on an American or Russian index.
    \item Relatedly, the difference between digital patriotism enacted in favour of Europe, versus specific countries within Europe. We observed commenters cite specific countries, but were not able to establish sufficient evidence that many people were making decisions on a patriotic basis more specific than Europe.
    \item Also related, the meaning of Europe in this context, its conflation with the EU (picked upon by some commenters as undesirable), or the UK's status as ``European'' (a notable corollary to investigate, given that the UK has its own, divergent position on regulation for digital sovereignty).
\end{itemize}

\section{Conclusion}

To an unprecedented degree, software adoption is being influenced by ideological concerns, in addition to instrumental ones. This paper has investigated one such concern: that of digital patriotism. Through two studies, we have seen how the ideology of digital sovereignty, which emerged from a set of geopolitical events starting in the 2010s, and which since 2020 has become official policy for States across Europe, translates into the motivation to adopt or not adopt software on a consumer and business level. 

In the first study, an analysis of how and why governments switch away from  standards, we find that prior to 2020, the motivations to switch were largely related to costs and avoidance of lock-in, and the solutions proposed were the coordinated use of open-source software. However, starting in 2020, the cited motivations begin to refer to digital sovereignty, and the solutions proposed broaden to include direct investment in state-owned digital infrastructure and collaboration with the private sector.

In the second study, an analysis of comments on the online platform Hacker News describing software adoption rationales, we see how issues with de facto standards, negative sentiment towards American technology companies, and apprehension of the government of the USA, combine to motivate digital patriotism and the search specifically for European alternatives to de facto standards. However, there were challenges with this search: the available alternatives were often perceived as inferior, and there was skepticism about the benevolence of European companies. The willingness to accept and support an alternative even when it is perceived as functionally inferior is key to our formulation: digital patriotism is the adoption of a digital technology (viz., software) in support of digital sovereignty \emph{even if it entails compromise or sacrifice}.

Whether digital patriotism proves to be a lasting realignment or a transient response to geopolitical turbulence remains to be seen. What is clear is that software adoption can no longer be understood as apolitical. The software stack embodies a value stack: each dependency a declaration, each update a renewal of allegiance. The install button is becoming a site of ideological contest: a raised flag.

\section*{AI Disclosure Statement}
Some passages in Section~\ref{sec:background}, initially written manually by the authors, were processed using a language model in order to make them briefer. Excerpts from non-English materials were translated into English using a language model. The accessibility description for Figure 1 was generated with the help of a language model. All language model output was reviewed and edited as appropriate by the authors.

\bibliographystyle{ACM-Reference-Format}
\bibliography{references, secondary-references}

\appendix
\section{Sources for Switching Events}
\label{apx:government-switches}

All sources in this appendix were last accessed 3 January 2026.

\begin{table}[!h]
\centering
\begin{tabular}{|l|p{0.75\textwidth}|}
\hline
\textbf{Source Group ID} & \textbf{Source URLs} \\ \hline

S-1 &
\parbox[t]{0.75\textwidth}{
No primary source available\\
Secondary sources:\\
\url{https://www.zdnet.fr/actualites/la-gendarmerie-nationale-passe-a-openoffice-39203431.htm}\\
\url{https://en.wikipedia.org/wiki/GendBuntu}
}
\\ \hline

S-2 &
\parbox[t]{0.75\textwidth}{
No primary source available\\
Secondary sources:\\
\url{https://www.techmonitor.ai/technology/french_national_assembly_confirms_move_to_open_source}
}
\\ \hline

S-3 &
\parbox[t]{0.75\textwidth}{
Primary source:\\
\url{https://interoperable-europe.ec.europa.eu/sites/default/files/document/2011-12/IDABC.OSOR.casestudy.Gendarmerie.10.pdf}\\
Secondary sources:\\
\url{https://arstechnica.com/information-technology/2009/03/french-police-saves-millions-of-euros-by-adopting-ubuntu/}\\
\url{https://en.wikipedia.org/wiki/GendBuntu}
}
\\ \hline

S-4 &
\parbox[t]{0.75\textwidth}{
No primary source available\\
Secondary sources:\\
\url{https://interoperable-europe.ec.europa.eu/collection/ehealth/news/dk-25000-hospital-staff-cop}
}
\\ \hline

S-5 &
\parbox[t]{0.75\textwidth}{
No primary source available\\
Secondary sources:\\
\url{https://interoperable-europe.ec.europa.eu/collection/open-source-observatory-osor/news/leipzig-switching-open}
}
\\ \hline

S-6 &
\parbox[t]{0.75\textwidth}{
No primary source available\\
Secondary sources:\\
\url{https://interoperable-europe.ec.europa.eu/collection/open-source-observatory-osor/document/toulouse-saves-1-million-euro-libreoffice}
}
\\ \hline

S-7 &
\parbox[t]{0.75\textwidth}{
Primary source:\\
\url{https://www.greeklug.gr/images/stories/drastiriotites/pressrelease-20120327-dim-libreoffice.pdf}\\
Secondary sources:\\
\url{https://www.greeklug.gr/index.php?option=com_content&view=article&id=204\%3Apressrelease-20120327-dim-libreoffice&catid=1\%3Aannouncementsnews&Itemid=65&lang=el}\\
\url{https://interoperable-europe.ec.europa.eu/collection/open-source-observatory-osor/news/greek-municipality-pilea-h}
}
\\ \hline

S-8 &
\parbox[t]{0.75\textwidth}{
No primary source available\\
Secondary sources:\\
\url{https://www.zdnet.com/article/libreoffice-love-sees-microsoft-removed-from-thousands-of-pcs-in-south-tyrol/}\\
\url{https://www.zdnet.com/article/ditch-microsoft-office-or-take-a-pay-cut-which-would-you-choose/}
}
\\ \hline

\end{tabular}
\end{table}

\begin{table}[!h]
\centering
\begin{tabular}{|l|p{0.75\textwidth}|}
\hline
\textbf{Source Group ID} & \textbf{Source URLs} \\ \hline

S-9 &
\parbox[t]{0.75\textwidth}{
Primary source:\\
\url{https://dgtic.gva.es/va/-/la-generalitat-implanta-software-libre-ofimatico-en-todos-los-ordenador-1}\\
Secondary sources:\\
\url{https://www.dcdata.co.za/valencia-region-government-completes-switch-to-libreoffice/}\\
\url{https://interoperable-europe.ec.europa.eu/collection/open-source-observatory-osor/news/valencia-region-government-co}
}
\\ \hline

S-10 &
\parbox[t]{0.75\textwidth}{
No primary source available\\
Secondary sources:\\
\url{https://www.zdnet.com/article/like-driving-a-ferrari-at-20mph-why-one-region-ditched-microsoft-office-for-libreoffice/}
}
\\ \hline

S-11 &
\parbox[t]{0.75\textwidth}{
Primary source:\\
\url{https://www.gov.uk/government/news/open-document-formats-selected-to-meet-user-needs}\\
Secondary sources:\\
\url{https://www.zdnet.com/article/uk-makes-odf-its-official-documents-format-standard/}
}
\\ \hline

S-12 &
\parbox[t]{0.75\textwidth}{
Primary source:\\
\url{https://www.libreitalia.org/accordo-di-collaborazione-tra-libreitalia-e-difesa/}\\
Secondary sources:\\
\url{https://www.zdnet.com/article/italian-ministry-of-defense-moves-to-libreoffice/}\\
\url{https://www.zdnet.com/article/from-microsoft-to-libreoffice-how-italys-military-is-starting-its-march-to-open-source/}
}
\\ \hline

S-13 &
\parbox[t]{0.75\textwidth}{
Primary source:\\
\url{https://concetticontrastivi.org/wp-content/uploads/2024/04/barcelona_data_management_0.1.en_.pdf}\\
Secondary sources:\\
\url{https://ajuntamentdebarcelona.github.io/ethical-digital-standards-site/tech-sovereignty/0.1/general-principles.html}\\
\url{https://ajuntamentdebarcelona.github.io/ethical-digital-standards-site/free-soft/0.2/introduction.html}
}
\\ \hline

S-14 &
\parbox[t]{0.75\textwidth}{
No primary source available\\
Secondary sources:\\
\url{https://interoperable-europe.ec.europa.eu/collection/open-source-observatory-osor/news/public-money-public-code}\\
\url{https://www.linuxinsider.com/story/Barcelona-Opts-for-Breath-of-Open-Source-Fresh-Air-85077.html}\\
\url{https://www.techradar.com/news/barcelona-abandons-windows-and-office-goes-with-linux-instead}\\
\url{https://www.digitaltrends.com/computing/barcelona-dumps-microsoft-windows-linux/}
}
\\ \hline

S-15 &
\parbox[t]{0.75\textwidth}{
No primary source available\\
Secondary sources:\\
\url{https://blog.documentfoundation.org/blog/2018/11/22/municipality-of-tirana/}
}
\\ \hline

S-16 &
\parbox[t]{0.75\textwidth}{
Primary source:\\
\url{https://www.landtag.ltsh.de/infothek/wahl19/drucks/02000/drucksache-19-02056.pdf}\\
No secondary sources.
}
\\ \hline

\end{tabular}
\end{table}

\begin{table}[!h]
\centering
\begin{tabular}{|l|p{0.75\textwidth}|}
\hline
\textbf{Source Group ID} & \textbf{Source URLs} \\ \hline

S-17 &
\parbox[t]{0.75\textwidth}{
Primary source:\\
\url{https://rathaus.dortmund.de/dosys/doRat.nsf/NiederschriftXP.xsp?action=openDocument&documentId=28B18161663E54D5C12586A8002F14EA}\\
Secondary sources:\\
\url{https://fsfe.org/news/2021/news-20210331-01.html}\\
\url{https://blog.do-foss.de/do-foss/}
}
\\ \hline

S-18 &
\parbox[t]{0.75\textwidth}{
Primary source:\\
\url{https://www.schleswig-holstein.de/DE/landesregierung/ministerien-behoerden/I/Presse/PI/2024/CdS/240403_cds_it-arbeitsplatz}\\
Secondary sources:\\
\url{https://euro-stack.com/blog/2025/3/schleswig-holstein-open-source-digital-sovereignty}\\
\url{https://blog.documentfoundation.org/blog/2024/04/04/german-state-moving-30000-pcs-to-libreoffice/}\\
\url{https://blog.documentfoundation.org/blog/2025/03/13/updates-on-schleswig-holstein-moving-to-libreoffice/}\\
\url{https://itsfoss.com/news/german-state-ditches-microsoft/}\\
\url{https://www.france24.com/en/live-news/20250613-we-re-done-with-teams-german-state-hits-uninstall-on-microsoft}\\
\url{https://www.heise.de/news/Schleswig-Holsteins-Digitalminister-Albrecht-ueber-den-Wechsel-zu-Open-Source-6221361.html}\\
\url{https://www.theregister.com/2024/04/04/germanys_northernmost_state_ditches_windows/}\\
\url{https://itsfoss.com/news/german-state-foss/}\\
\url{https://www.heise.de/news/Open-Source-vor-Schleswig-Holstein-will-sich-vollstaendig-von-Microsoft-loesen-4079834.html}
}
\\ \hline

S-19 &
\parbox[t]{0.75\textwidth}{
Primary source:\\
\url{https://www.tweedekamer.nl/debat_en_vergadering/plenaire_vergaderingen/details/activiteit?id=2025A02077}\\
Secondary sources:\\
\url{https://www.theregister.com/2025/03/19/dutch_parliament_us_tech/}
}
\\ \hline

S-20 &
\parbox[t]{0.75\textwidth}{
Primary source:\\
\url{https://politiken.dk/viden/tech/art10411353/Nu-er-det-nok-K\%C3\%B8benhavn-og-Aarhus-vil-g\%C3\%B8re-sig-uafh\%C3\%A6ngige-af-Microsoft}\\
Secondary sources:\\
\url{https://archive.is/bL9FV}\\
\url{https://world.hey.com/dhh/denmark-gets-more-serious-about-digital-sovereignty-7736f756}
}
\\ \hline

S-21 &
\parbox[t]{0.75\textwidth}{
Primary source:\\
\url{https://politiken.dk/viden/tech/art10437680/Caroline-Stage-udfaser-Microsoft-i-Digitaliseringsministeriet}\\
Secondary sources:\\
\url{https://itsfoss.com/news/denmark-set-to-replace-microsoft/}\\
\url{https://blog.documentfoundation.org/blog/2025/07/08/danish-ministry-switching-from-microsoft-office-365-to-libreoffice/}\\
\url{https://www.theregister.com/2025/06/13/danish_department_dump_microsoft/}\\
\url{https://nordjyske.dk/nyheder/politik/digitaliseringsminister-vil-udfase-microsoft-i-sit-eget-ministerium/5616096}\\
\url{https://vezirajans.com/global-shift-countries-moving-away-from-windows-and-office-to-embrace-linux}
}
\\ \hline

\end{tabular}
\end{table}

\begin{table}[!h]
\centering
\begin{tabular}{|l|p{0.75\textwidth}|}
\hline
\textbf{Source Group ID} & \textbf{Source URLs} \\ \hline

S-22 &
\parbox[t]{0.75\textwidth}{
Primary source:\\
\url{https://www.lyon.fr/actualite/action-municipale/la-ville-de-lyon-renforce-sa-souverainete-numerique}\\
See also:\\
\url{https://territoirenumeriqueouvert.fr/le-projet-tno/}\\
Secondary sources:\\
\url{https://www.zdnet.com/article/this-city-is-dumping-microsoft-office-and-windows-for-onlyoffice-and-linux-heres-why/}\\
\url{https://www.techradar.com/pro/one-of-frances-largest-cities-has-now-also-ditched-microsoft-for-open-source-software}\\
\url{https://interoperable-europe.ec.europa.eu/collection/open-source-observatory-osor/news/municipality-lyon-moves-towards-open-source}
}
\\ \hline

S-23 &
\parbox[t]{0.75\textwidth}{
Sources documenting attempted switches and reversion in Munich:\\
2003, switch to Linux: \url{https://www.zdnet.com/article/munich-breaks-with-windows-for-linux/}\\
2004, migration put on hold: \url{https://www.zdnet.com/article/munichs-linux-plans-attract-international-attention/}\\
2005, further delays: \url{https://www.zdnet.com/article/reality-bytes-as-munich-delays-linux-project/}\\
2017, Munich city council votes to return to de facto standard: \url{https://www.heise.de/news/Endgueltiges-Aus-fuer-LiMux-Muenchener-Stadtrat-setzt-den-Pinguin-vor-die-Tuer-3900439.html}\\
2020, a new coalition government agrees to switch again to OSS software: \url{https://www.zdnet.com/article/linux-not-windows-why-munich-is-shifting-back-from-microsoft-to-open-source-again/}\\
2023, Munich again returns to de facto standard: \url{https://itsfoss.com/munich-linux-failure/}

}
\\ \hline

S-24 &
\parbox[t]{0.75\textwidth}{
Sources documenting reversed switch in Lower Saxony police and tax administration:\\
\url{https://www.heise.de/news/Neues-Computersystem-der-Polizei-Niedersachsen-faellt-oft-aus-88803.html}\\
\url{https://www.heise.de/news/Polizei-Niedersachsen-will-von-Linux-zurueck-zu-Microsoft-2440829.html}\\
\url{https://www.heise.de/news/Linux-Aus-Niedersachsen-will-knapp-13-000-Rechner-auf-Windows-umstellen-4119380.html}\\
\url{https://www.theregister.com/2018/07/27/lower_saxony_to_dump_linux/}
}

\\ \hline

S-25 &
\parbox[t]{0.75\textwidth}{
Sources documenting failed migration in Vienna:\\
\url{https://www.theregister.com/2005/11/22/vienna_goes_open/}\\
\url{http://freesoftwaremagazine.com/articles/vienna_failed_to_migrate_to_linux_why/}
}
\\ \hline

\end{tabular}
\end{table}

\section{Hacker News Threads Analysed}
\label{apx:hn-threads}

The analysis was conducted on a snapshot of the following Hacker News threads taken on 14 June 2025.

\begin{enumerate}

\item \url{https://news.ycombinator.com/item?id=43458509}

\item \url{https://news.ycombinator.com/item?id=29627097}

\item \url{https://news.ycombinator.com/item?id=40695754}

\item \url{https://news.ycombinator.com/item?id=43329992}

\item \url{https://news.ycombinator.com/item?id=43104684}

\item \url{https://news.ycombinator.com/item?id=5993947}

\item \url{https://news.ycombinator.com/item?id=43589183}

\item \url{https://news.ycombinator.com/item?id=42932956}

\item \url{https://news.ycombinator.com/item?id=43548032}

\item \url{https://news.ycombinator.com/item?id=43702381}

\item \url{https://news.ycombinator.com/item?id=30853025}

\item \url{https://news.ycombinator.com/item?id=22963328}

\item \url{https://news.ycombinator.com/item?id=40406872}

\item \url{https://news.ycombinator.com/item?id=35466273}

\item \url{https://news.ycombinator.com/item?id=5861717}

\item \url{https://news.ycombinator.com/item?id=37404129}

\item \url{https://news.ycombinator.com/item?id=42913477}

\item \url{https://news.ycombinator.com/item?id=43113054}

\item \url{https://news.ycombinator.com/item?id=42820723}

\item \url{https://news.ycombinator.com/item?id=25006716}

\item \url{https://news.ycombinator.com/item?id=43318738}

\item \url{https://news.ycombinator.com/item?id=29577039}

\item \url{https://news.ycombinator.com/item?id=42944752}

\item \url{https://news.ycombinator.com/item?id=42874533}

\item \url{https://news.ycombinator.com/item?id=6001704}

\item \url{https://news.ycombinator.com/item?id=41199627}

\item \url{https://news.ycombinator.com/item?id=39653805}

\item \url{https://news.ycombinator.com/item?id=43342167}

\item \url{https://news.ycombinator.com/item?id=2811748}

\item \url{https://news.ycombinator.com/item?id=39134479}

\item \url{https://news.ycombinator.com/item?id=43353953}

\item \url{https://news.ycombinator.com/item?id=39223257}

\item \url{https://news.ycombinator.com/item?id=39141716}

\item \url{https://news.ycombinator.com/item?id=39615951}

\item \url{https://news.ycombinator.com/item?id=42915980}

\item \url{https://news.ycombinator.com/item?id=43268988}

\end{enumerate}

\section{Shift in Legislative Focus}
\label{apx:legislative-shift}

As noted in Section~\ref{sec:study1}, government motivations for switching away from de facto standards, as well as the solutions they choose instead, shifted significantly around 2020. The motivations shifted from the instrumental rationale of cost savings and avoiding lock-in towards the value rationale in the emerging ideology of digital sovereignty. Concomitantly, the proposed solutions also shifted from a focus on coordinated use of open-source software to direct investment, not only in open-source but also in the domestic private sector. 

In studying these switches, we came across a similar general shift in government efforts for legislation and investment that we document in this section, to complement the analysis in Section~\ref{sec:study1}. The objective in this section is to help explain the broader regulatory context within which the shift in rationales occured, not to provide an exhaustive analysis.

This shift is clearly exemplified in the series of Acts of the European Parliament that culminate in the Interoperable Europe Act of 2024 \cite{eu_2024_interoperable_europe_act}. Its predecessors are the interoperability solutions for European public administrations (ISA) effective 2009-2015 \cite{eu_2009_isa_decision}, and its update, the ISA2, effective from 2016-2021 \cite{eu_2015_isa2_decision}. While the 2024 Act still retains the vestiges of the original aims of interoperability, its aims and mechanisms have shifted significantly. The 2024 Act contains language of sovereignty (\textit{``boost the global competitiveness, resilience and open strategic autonomy of the Union''}) and an increased willingness to collaborate with the private sector (\textit{``cooperate with innovative organisations, including companies and not-for-profit entities, in design, development and operation of public services''}; \textit{``tap in particular into the Union’s rich reservoir of technology start-ups and SMEs''}), whereas the original ISA was aimed simply at facilitating \textit{``the efficient and effective electronic cross-border and cross-sectoral interaction between European public administrations''}.

Early projects aimed at improving interoperability between government services and public data, such as the European IDABC programme (2005-2009) \cite{eurlex_idabc_summary} and the French Etalab (founded 2011) \cite{wikipedia_etalab}. Public licenses were developed to share software and data, such as the European Union Public License (2007) \cite{eupl_1_1_2007}, the UK Government Open License (2010) \cite{ukgov_2015_public_data_barrier}, and the French Open License (2011) \cite{wikipedia_licence_ouverte}. From 2006-2010 the European Commission funded the Open Source Observatory and Repository to share expertise on OSS across public admininstrations \cite{interoperableeurope2012_osor_knowing_sharing}. The European Commission published open source strategies for 2014-2017 and 2020-2023 \cite{ec2020_oss_strategy}, part of which involved starting a repository for open source software \cite{ec2021_open_source_licensing_decision, codeeuropa2025_readme}.

However, as mentioned previously, the emphasis began to shift in 2020. For instance, in 2020 the city of Munich reaffirmed its commitment to open source, but included mention of digital sovereignty (\textit{``digital sovereignty means that central and critical areas of digital public life remain in the hands of citizens or municipalities''} [translated]) \cite{muenchen2020_koalitionsvereinbarung}. In 2022 the German Federal Ministry for Economic Affairs and Energy began funding the Sovereign Tech Fund (now the Sovereign Tech Agency) to fund the development of open source software with the aim of strengthening digital sovereignty (\textit{``Digital sovereignty empowers us to shape our own future. This requires more than simply using digital technologies; we need to actively be part of building them''}) \cite{sovereigntechagency2025_website}. In 2022 the German Federal Ministry of the Interior greenlighted the foundation of ZenDis, a state-owned enterprise that develops a software repository and online office suite, with the aim of addressing \textit{``a serious loss of control over state IT [...] With a comprehensive sovereignty package consisting of a platform, products and consulting''} \cite{zendis2025_homepage}.

\subsection{Artificial Intelligence}
The intensity of efforts to collaborate with the private sector and the increasing appetite for government investment in infrastructure to pursue the aims of digital sovereignty are particularly acute in the case of Artificial Intelligence (AI). AI is favoured on multiple accounts: it is the technology \emph{du jour} and thus a symbol of the frontier of digital technology and the sociotechnical imaginary of the ``arms race'', and it is a technology that appears to readily translate capital investment to results on several fronts, from talent, to hardware infrastructure (datacenters), to software (training and inference of large foundation models). In essence, unlike with prior recent frontier technologies (cryptocurrency, mixed reality and ``metaverse'', etc.), the obtainable result is perceived to be in direct proportion to the magnitude of investment.

For example, in 2023 the Dutch RVO (an agency of the Ministry of Economic Affairs) began to fund development of a sovereign language model dubbed GPT-NL, which aims to be \textit{``a language model for the Dutch language and culture: reliable, transparent, reciprocal, and sovereign''} (from Dutch: \textit{``een taalmodel voor de Nederlandse taal en cultuur: betrouwbaar, transparant, wederkerig, en soeverein''}) \cite{gptnl2025_verantwoord_alternatief}. It is not alone -- Politico documents several similar efforts across Europe, including in Denmark, Sweden, Finland, Latvia, France, Poland, Austria, Italy, Romania, Bulgaria, Greece, Spain, and Portugal -- almost universally as public or public-private partnerships \cite{volpicelli_coi_2024_eu_chatbots}. In 2025 an open language model was announced by ETH Zurich, EPFL, and the Swiss National Supercomputing Centre in collaboration with NVIDIA and HPE/Cray, \textit{``as an enabler of sovereign AI''} \cite{meyer_anchisi_2025_public_good}. The OpenEuroLLM project is a public-private collaboration between multiple universities and companies \textit{``to develop next-generation open-source language models [...] to improve Europe’s competitiveness and digital sovereignty''}, launched in 2025 \cite{openeurollm2025_launch_press}.

These are supported by broader funding and incubation efforts aimed at stimulating AI development in Europe. For instance, the French Alliance incubator (styled ``ALLiaNCE''), launched in 2024, aims \textit{``to promote the adoption of AI in the State in an ethical and sovereign manner''} (from French: \textit{``pour favoriser l'adoption de l'IA dans l'État de manière éthique et souveraine''}) \cite{dinum2024_alliance_incubator}. The European AI office was established in 2024 to harmonise governance of AI in the Union and \textit{``to foster the development, use and uptake of artificial intelligence [... that] meets a high level of protection of public interests''} \cite{ec2024_ai_office_decision}, and in 2025 began leading the implementation of the AI Continent Action Plan outlining commitments to make Europe a leader in computing infrastructure, data, algorithms, talent, and compliance \cite{ec2025_ai_continent_action_plan} and the Apply AI strategy which \textit{``promotes a `buy European' approach [...] to increase EU’s technological sovereignty''} \cite{ec2025_apply_ai_strategy}. In 2024 the European Innovation Council announced investments of 1.4 billion EUR in ``deep tech'' including generative AI \cite{ec2024_eic_invest_14b}, as part of the Strategic Technologies for Europe Platform (STEP) regulation adopted earlier in the year, which was motivated by the observation that \textit{``immediate action is required to support the development and manufacturing in the Union of critical technologies, which constitute the Union’s primary strategic deficiencies [...] thereby reducing the Union’s strategic dependencies''} \cite{eu2024_step}. In 2025, the EU launched the DARE project, allocating 120 million EUR to develop high performance computing hardware for AI for ``advancing European sovereignty'' \cite{eurohpcju_2025_riscv_dare}, and the InvestAI initiative, allocating 200 billion EUR for investment in AI, including 20 billion EUR for European AI ``gigafactories'' \cite{ec2025_investai_press}.

In 2025, the European Commission's joint white paper for European defense readiness 2030 explicitly connects technology and digital sovereignty with military, security and defense: \textit{``Geopolitical rivalries have not only led to a new arms race but have also provoked a global technology race. Technology will be the main feature of competition in the new geopolitical environment. A handful of critical and foundational technologies like AI, quantum, biotech, robotics, and hypersonic are key inputs for both long term economic growth, and military pre-eminence''} \cite{eu_commission_2025_white_paper_defence}. Later that year, the EU council State of the Digital Decade report, subtitled \textit{``keep building the EU's sovereignty and digital future''}, makes a similar explicit connection \cite{council2025_st10407_init}.

Similar efforts are underway in the United Kingdom. In 2025 the UK Department for Science, Innovation and Technology published the AI Opportunities Action Plan, in which it sets the aim for \textit{``Britain to step up; to shape the AI revolution rather than wait to see how it shapes us''} \cite{clifford2025_ai_opportunities}. This involved the creation of the Sovereign AI Unit, with up to 500 million GBP of funding, to \textit{``enhance UK national security''} \cite{dsit2025_sovereign_ai_unit}, and was followed shortly afterwards by the AI Growth Lab, a regulatory sandbox scheme aimed at accelerating innovation in AI \cite{dsit2025_ai_regulation_blueprint}. Similarly in 2025 UK Research and Innovation offered a 1.6 million GBP grant for \textit{``AI technologies with state of the art performance [...] to support the UK’s AI sovereignty objectives''} \cite{innovateuk2025_sovereign_ai_poc}.

\end{document}